\documentclass[prevtex,nofootinbib,12pt]{revtex4} 

\usepackage{amsmath,amscd}

\begin{document}

\title{A novel isospectral deformation chain in supersymmetric quantum mechanics}

\author{Bj\o rn Jensen}
\address{Faculty of Micro and Nano Systems Technology, Vestfold University College, N-3103 T\o nsberg Norway}

\email{bjorn.jensen@hive.no}

\date{\today}


\begin{abstract}
We present a novel isospectral deformation chain in supersymmetric quantum mechanics.  The chain is applied to the Coulomb potential.  
\end{abstract}

\maketitle

\section{Introduction} 

In \cite{Jensen} we introduced within the framework of supersymmetric quantum mechanics an isospectral deformation on the form
\begin{equation}
W(x)\rightarrow\hat{W}_0(x)=F_0(x)W(x)\, ,
\end{equation}
where $W(x)$ is some known superpotential and $F_0(x)$ some function to be determined by the isospectrality condition
\begin{equation}
\hat{V}_+(x)\equiv \hat{W}_0^2(x)+\hat{W}_0'(x)=W^2(x)+W'(x)\equiv V_+(x)\, .
\end{equation} 
It was shown that Eq.(1) includes the only previously explored deformation of this kind, which has the form (\cite{Asim}, e.g.)
\begin{equation}
W(x)\rightarrow\hat{W}_0(x)=W(x)+f(x)\, .
\end{equation}
$f(x)$ is some function which is determined by Eq.(2). In this paper we expand the deformation Eq.(1) in various directions and study the implications drawn from the isospectrality condition.  We show in particular that the deformation Eq.(1) is the root of an infinitely long and recursively  generated chain of deformations.

In the next section we briefly review some findings in \cite{Jensen}.  We then define and derive an explicit recursive scheme for generating novel isospectral deformations.  We then discuss possible generalizations of this scheme. We next apply the recursive scheme to the Coulomb potential. We conclude and discuss some of our findings in the last section.

\section{Base deformations}

The deformation Eq.(1) implies the following differential equation for $F_0(x)$ \cite{Jensen}\footnote{We will often rewrite fractions on the form $W'(x)/W(x)$ as the logarithmic derivative of $W(x)$ as a formal tool.  Caution must of course be exercised when using the corresponding expressions in actual computations.}
\begin{eqnarray}
\frac{d}{dx}F_0(x)&+&(\frac{d}{dx}\ln W(x))F_0(x)+W(x)F_0^2(x)= W(x)+\frac{d}{dx}\ln W(x)\, .
\end{eqnarray}
This is the generalized Riccati equation \cite{Matte} . If one particular solution $F_{00}(x)$ of Eq.(4) is known another solution is given by \cite{Boyce}
\begin{equation}
F_0(x)=F_{00}(x)+\frac{1}{X_0(x)}\, ,
\end{equation}
where $X_0(x)$ solves the equation
\begin{equation}
\frac{d}{dx}X_0(x)-(\frac{d}{dx}\ln W(x)+2F_{00}(x)W(x))X_0(x)=W(x)\, .
\end{equation}
Eq.(6) can be solved by elementary means. The resulting superpotential $\hat{W}_0(x)$ is given by \cite{Jensen}
\begin{eqnarray}
\hat{W}_0(x)&=&(F_{00}(x)+\frac{1}{X_0(x)})W(x)\equiv \hat{W}_{00}(x)+\frac{1}{X_0(x)}W(x)=\nonumber\\
&=&F_{00}(x)W(x)+\frac{e^{-2\int^x F_{00}(t)W(t)dt}}{C_{01} +\int^x e^{-2\int^u F_{00}(t)W(t)dt}du}\, .
\end{eqnarray}
$C_{01}$ is an integration constant, which we will assume to be real. We have explicitly introduced upper integration limits in Eq.(7) in order to avoid sign ambiguities. This explains the difference in the sign in the denominator in Eq.(7) compared with (2.5) in \cite{Jensen} where the reverse order of integration in one of the integrals was implicitly assumed. We do not specify the lower integration limits in Eq.(7).  These are not important, of course, since the values of the integrals there can essentially be absorbed into $C_{01}$. When $F_{00}(x)=1$ we identically rederive Eq.(3) and the corresponding expression discussed in \cite{Asim}. We can by simple inspection see that the particular solution $F_0(x)=1$, the {\it identity deformation}, solves Eq.(4). This deformation corresponds to the limit $C_{01}\rightarrow\infty$ in Eq.(7) with $F_{00}(x)=1$. In the limit $C_{01}\rightarrow\infty$ we generally get $\hat{W}_0(x)=\hat{W}_{00}(x)$. This deformation will play a pivotal role in this work; it will represent the base of a recursive scheme for generating novel isospectral deformations. We will therefore refer to a particular $\hat{W}_{00}(x)$ as {\it a base deformation} in the following. 

In order to expand the space of concrete isospectral deformations further we transform Eq.(4) into an ordinary second order differential equation by the substitution
\begin{equation}
F_0(x)=\frac{1}{W(x)}\frac{d}{dx}\ln U_0(x)\, .
\end{equation}
This substitution gives rise to the following linear homogeneous equation
\begin{equation}
-\frac{d^2}{dx^2}U_0(x)+V_+(x)U_0(x)=0\, .
\end{equation}
The special solution $F_0(x)=1$ is generated by the solution
\begin{equation}
U_0(x)\sim e^{\int^x W(t)dt}\, .
\end{equation}
The particular solutions for $F_0(x)$ stemming from these equations can be fed into Eq.(7) (as $F_{00}(x)$) and thus expand the space of available concrete deformations. The physical potential $\hat{V}_-(x)$ generated by $\hat{W}_0(x)$ can in general thus be written \cite{Jensen}\footnote{Note that the corresponding expression in \cite{Jensen} ((2.14)) is misprinted.}
\begin{eqnarray}
\hat{V}_-(x)&\equiv& \hat{W}_0^2(x)-\hat{W}_0'(x)=\hat{W}_{00}^2(x)-\hat{W}_{00}'(x)+\nonumber\\
&+&\frac{4\hat{W}_{00}(x)e^{-2\int^x\hat{W}_{00}(t)dt}}{C_{01}+\int^xe^{-2\int^u\hat{W}_{00}(t)dt}du}+2\left[ \frac{e^{-2\int^x\hat{W}_{00}(t)dt}}{C_{01}+\int^xe^{-2\int^u\hat{W}_{00}(t)dt}du}   \right]^2
\end{eqnarray}
with
\begin{equation}
\hat{W}_{00}(x)=\frac{d}{dx}\ln U_0(x)\, .
\end{equation}

\section{Recursive linear deformations}

Although the Riccati equation can be transformed into an ordinary second order differential equation the non-linearity of the equation allows for a solution space which is larger than the one associated with linear differential equations of second order, as became evident in the previous section. It is therefore natural to ask whether the non-linearity of the Riccati equation implies even more isospectral deformations than the ones we already have deduced \cite{Jensen}.  We will explore this question in this and the next section.  

\subsection{The sum}

Let us entertain the following idea. Assume that we have derived a particular base deformation $\hat{W}_{00}(x)$ from an explicitly given superpotential $W(x)$. Then assume that we {\it add} another term $F_1(x)W(x)$ to that deformation such that we in principle get a novel deformation on the form $\hat{W}(x)=F_{10}(x)W(x)+\hat{W}_{00}(x)$. After determining $F_{10}(x)$ from the isospectrality condition Eq.(2) add yet another term of this kind to the deformation. Let us assume that this process can be repeated indefinitely. Will terms added in this manner give rise to novel deformations? We will in the following show that they do.  This represents a recursive deformation scheme.  

Following the basic idea, after $m$ iterations we thus have the general recursive linear (in $W(x)$) deformation
\begin{equation}
\hat{W}_{m0}(x)=(\sum_{i=0}^{m}\lambda_iF_{i0}(x))W(x)=\lambda_mF_{m0}(x)W(x)+\hat{W}_{(m-1)0}(x)\,\, ,\,\, \lambda_0\equiv 1\, .
\end{equation}
The $\lambda_i$'s are assumed to be independent real constants.  Starting with a known superpotential $m$ consecutive applications of the isospectrality condition yields the following set of equations 
\begin{equation}
\left\{\begin{array}{l}
F_{00}'(x)+[\ln W(x)]'F_{00}(x)+W(x)F_{00}^2(x)=W(x)+(\ln W(x))'\, ,\\
 F_{10}'(x)+[(\ln W(x))'+2F_{00}(x)W(x)]F_{10}(x)+\lambda_1W(x)F_{10}^2(x)=0\, ,\\
 F_{20}'(x)+[(\ln W(x))'+2(F_{00}(x)+\lambda_1F_{10}(x))W(x)]F_{20}(x)+\lambda_2W(x)F_{20}^2(x)=0\, ,\\
 \vdots\\
 \vdots\\
 F_{m0}'(x)+[(\ln W(x))'+2\hat{W}_{(m-1)0}(x))]F_{m0}(x)+\lambda_mW(x)F_{m0}^2(x)=0\, .
\end{array}
\right.
\end{equation}
The first equation in Eq.(14) coincides of course per definition with Eq.(4). Note that $F_{j0}(x)=1$ only solves the first equation in Eq.(14). Let us consider an arbitrary iteration level $n\, (\neq 0)$ and make the following substitution in Eq.(14)
\begin{equation}
F_{n0}(x)=\frac{1}{W(x)}(\ln U_n(x))'\, .
\end{equation}
The equation for $F_n(x)$ can then be written
\begin{equation}
U''_n(x)+(\lambda_n-1)\frac{[U'_n(x)]^2}{U_n(x)}+2\hat{W}_{(n-1)0}(x)U'_n(x)=0\, .
\end{equation}
This equation corresponds to Eq.(9) in the case when $n=0$. It reduces in general to an ordinary linear differential equation only when $\lambda_n=1\, ,\, \forall n\neq 0$.  We will focus on this special case in this paper.   

The general solution of Eq.(16) for arbitrary $n\neq 0$, and with $\lambda_n$ set to unity, can be found by elementary means, and we deduce that
\begin{eqnarray}
F_{n0}(x)W(x)&=&\frac{C_{n2}e^{-2\int^x \hat{W}_{(n-1)0}(t)dt}}{C_{n1}+C_{n2}\int^x e^{-2\int^u \hat{W}_{(n-1)0}(t)dt}du}=\nonumber\\
&=&\frac{d}{dx}\ln (C_{n1}+C_{n2}\int^x e^{-2\int^u \hat{W}_{(n-1)0}(t)dt}du)\, .
\end{eqnarray}
$C_{n1}$ and $C_{n2}$ are integration constants, which we assume to be real. We can reduce the number of integration constants to one at each iteration level, but we will stick to the habit of explicitly writing down the actual number of constants in order to make it easier to compare the various formulas we deduce, which stem from both second and first order differential equations. We also note that the structure of $F_{n0}(x)$ implies that previous deformations are not regenerated in general. Of course, this does not exclude this possibility to arise, as we will see in Section 5. Hence, $m$ in Eq.(13) has in principle no natural upper bound. From Eq.(13) and Eq.(17) we get the following expression for the superpotential at iteration level $m$
\begin{eqnarray}
\hat{W}_{m0}(x)&=&\hat{W}_{00}(x)+\sum_{j=1}^{m}\frac{d}{dx}\ln (C_{j1}+C_{j2}\int^x e^{-2\int^u \hat{W}_{(j-1)0}(t)dt}du)=\nonumber\\
&=&\hat{W}_{00}(x)+\frac{d}{dx}\ln\prod_{j=1}^{m}(C_{j1}+C_{j2}\int^x e^{-2\int^u \hat{W}_{(j-1)0}(t)dt}du)\equiv\nonumber\\
&\equiv&\hat{W}_{00}(x)+\frac{d}{dx}\ln P_m(x)\, .
\end{eqnarray}
From Eq.(18) we deduce that
\begin{equation}
e^{-2\int^x\hat{W}_{(j-1)0}(t)dt}=P_{j-1}^{-2}(x)e^{-2\int^x\hat{W}_{00}(t)dt}\,\, ;\,\, P_0^2(x)\equiv 1\, ,\, j\neq 0\, ,
\end{equation}
such that
\begin{equation}
\left\{\begin{array}{l}
P_1(x)=C_{11}+C_{12}\int^x e^{-2\int^u\hat{W}_{00}(t)dt}du\, ,\\
P_2(x)=P_1(x)(C_{21}+C_{22}\int^x P_1^{-2}(u)e^{-2\int^u\hat{W}_{00}(t)dt}du)\, ,\\
 \vdots\\
 \vdots\\
P_m(x)=P_{m-1}(x)(C_{m1}+C_{m2}\int^x P_{m-1}^{-2}(u)e^{-2\int^u\hat{W}_{00}(t)dt}du)\, .
\end{array}
\right.
\end{equation}
Hence,
\begin{eqnarray}
P_n(x)=\prod_{j=1}^n(C_{j1}+C_{j2}\int^x P_{j-1}^{-2}(u) e^{-2\int^u\hat{W}_{00}(t)dt}du)\, .
\end{eqnarray}
This last form of the $P_n(x)$ functions neatly exhibits how the base deformation $\hat{W}_{00}(x)$ generates the higher order deformations. Some of the details we have deduced so far in this subsection are presented in Figure 1.

\begin{figure}
\begin{equation}
\left\{
\begin{CD}
W @>\times F_0=F_{00} >> \hat{W}_{00} @> +F_{10}W >> \hat{W}_{10}=\hat{W}_{00}+(\ln P_1)' @> + F_{20}W >> \cdots @> + F_{m0}W >> \hat{W}_{m0}
\\
\\
W @>\times F_0=1>> \hat{W}_{00}=W @> +F_{10}W >> \hat{W}_{10}=W+(\ln P_1)' @> + F_{20}W >> \cdots @> + F_{m0}W >> \hat{W}_{m0}
\end{CD}
\right.\nonumber
\end{equation}
\caption{The upper line depicts the solvable deformation chain Eq.(13) to iteration level $m$. There is no upper bound on $m$. The $F_{j0}(x)$ functions are given in Eq.(17). The $P_j(x)$ functions are given in Eq.(18) and Eq.(20). They are functions of a base deformation $\hat{W}_{00}(x)$. A base deformation $\hat{W}_{00}(x)$ is generated by the zero-energy Schr\"odinger equation interacting with the partner potential $V_+(x)$, Eq.(9). The second line depicts the important special case when $F_0(x)=1$. This particular solution can be derived as a special case of Eq.(7) with $X_0^{-1}=0$, which can be achieved by $C_{01}\rightarrow\infty$, and $F_{00}(x)$ determined by Eq.(8) and the solution Eq.(10). Note that $\hat{W}_{10}(x)$ then coincides with Eq.(7) when $F_{00}(x)=1$ and $C_{01}$ is finite in Eq.(7); Eq.(7) is thus in this particular case regenerated by the scheme at the next recursion level, i.e.}
\end{figure}

Make the following substitution at each iteration level in Eq.(13)
\begin{equation}
F_{n0}(x)\rightarrow F_{n0}(x)+\frac{1}{X_n(x)}\, .
\end{equation}
This implies (with the $\lambda_m$'s reinstated in Eq.(13)) a generalized form $\hat{W}_{m}(x)$ of the superpotential $\hat{W}_{m0}(x)$
\begin{equation}
\hat{W}_{m0}(x)=\sum_{i=0}^m\lambda_iF_{i0}W(x)\Rightarrow\hat{W}_m(x)=\hat{W}_{m0}(x)+\sum_{i=0}^m\frac{\lambda_i}{X_i(x)}W(x)\, .
\end{equation}
From Eq.(14) we find that $X_n(x)$ satisfies the equation
\begin{equation}
\frac{d}{dx}X_n(x)-(\frac{d}{dx}\ln W(x)+2\hat{W}_{n0}(x))X_n(x)=\lambda_mW(x)\, .
\end{equation}
This equation is a generalization of Eq.(6). The $n$'th deformation term Eq.(17) thus changes into
\begin{eqnarray}
F_{n0}(x)W(x)\rightarrow F_{n0}(x)W(x)+\frac{d}{dx}\ln (C_{n3}+\lambda_n\int^x e^{-2\int^u\hat{W}_{n0}(t)dt}du)\, .
\end{eqnarray}
$C_{n3}$ are integrations constants, which we assume to be real. Eq.(25) implies that the more general expression for the superpotential in Eq.(23) can be written as
\begin{equation}
\hat{W}_m(x)=\hat{W}_{m0}(x)+\frac{d}{dx}\ln Q_m(x)\, ,
\end{equation}
where
\begin{equation}
Q_m(x)\equiv\prod_{i=0}^m (C_{i3}+\lambda_i\int^x e^{-2\int^u\hat{W}_{i0}(t)dt}du)\, .
\end{equation}
$m=0$ in Eq.(26) ($\lambda_0\equiv 1$) reproduces Eq.(7). In the special case when $\lambda_m=1\, ,\forall m$ in Eq.(13) we get
\begin{equation}
\hat{W}_m(x)=\hat{W}_{00}(x)+\frac{d}{dx}\ln P_m(x)+\frac{d}{dx}\ln Q_m(x)\, .
\end{equation}
When we compare the expressions for $(P_m(x))'$ and $(Q_m(x))'$ we find that they differ by just the last term in $(Q_m(x))'$.


\subsection{The product}

What happens if we in Eq.(13) assume a product structure instead of a sum structure ? Let us assume that we have determined a base deformation. Let this be the seed superpotential for the deformation
\begin{equation}
\hat{W}_{00}(x)\rightarrow\hat{W}_{10}(x)=F_{10}(x)\hat{W}_{00}(x)=F_{10}(x)F_{00}(x)W(x)\, ,
\end{equation}
where $F_{10}(x)$ is some function to be determined by the isospectrality condition. This product scheme can of course in principle be repeated an arbitrary number $m$ times 
\begin{equation}
\hat{W}_{m0}(x)=(\prod_{i=0}^m F_{i0}(x))W(x)=F_{m0}(x)\hat{W}_{(m-1)0}\, .
\end{equation}
This structure gives rise to the following set of equations
\begin{equation}
\left\{\begin{array}{l}
F_{00}'(x)+(\ln W(x))'F_{00}(x)+W(x)F_{00}^2(x)=W(x)+(\ln W(x))'\, ,\\
F_{10}'(x)+(\ln \hat{W}_{00}(x))'F_{10}(x)+\hat{W}_{00}(x)F_{10}^2(x)=\frac{1}{F_{00}(x)}(W+(\ln W(x))')\, ,\\
\vdots\\
\vdots\\
F_{m0}'(x)+(\ln\hat{W}_{(m-1)0}(x))'F_{m0}(x)+\hat{W}_{(m-1)0}(x)F_{m0}^2(x)=\frac{W(x)}{\hat{W}_{(m-1)0}(x)}(W(x)+(\ln W(x))')\, .
\end{array}
\right.
\end{equation}
Clearly, each iteration level depends on all the previous ones, and at each level we are dealing with a non-homogenous non-linear differential equation. Interestingly, by making the following substitution at an arbitrary iteration level $n\neq 0$
\begin{equation}
F_{n0}(x)=\frac{1}{\hat{W}_{(n-1)0}(x)}(\ln U_n(x))'\, ,
\end{equation}
where $U_n(x)$ is some function, the equations Eq.(31) all reduce to Eq.(9). Hence, attempting to generate novel deformations recursively via a product structure, of the kind above, fails. This conclusion was also reached in \cite{Jensen}, but at the level of the second order linear differential equation Eq.(9). 

\section{Recursive non-linear deformations}

We have so far only considered linear deformations. In this section we will briefly consider two non-linear deformation schemes.  Let us first  consider a polynomial kind of deformation.  That is, given a superpotential $\hat{W}_{(i-1)0}(x)$ which we will assume is derived, in some way or another, from some seed superpotential $W(x)$.  Consider then the polynomial deformation
\begin{equation}
\hat{W}_{i0}(x)=F_{i0}(x)W^k(x)+\hat{W}_{(i-1)0}(x)\, ;\, k\in \{1,2,3,\ldots\}\, .
\end{equation}
The isospectrality condition implies
\begin{equation}
F_{i0}'(x)+[k(\ln W(x))'+2\hat{W}_{(i-1)0}(x)]F_{i0}(x)+W^k(x)F_{i0}^2(x)=0\, .
\end{equation}
This is a Riccati type equation of the kind we have met earlier in this paper. Apparently, different $k$-values give rise to very different equations to solve. However, and rather intriguingly, all the possible $k$-values implies the same deformation.  This is seen by making the following substitution
\begin{equation}
F_{i0}(x)=\frac{1}{W^k(x)}\frac{U'_k(x)}{U_k(x)}\, ,
\end{equation}
where $U_k(x)$ is some function. This expression inserted into Eq.(34) gives
\begin{equation}
U''_k(x)+2\hat{W}_{i-1}(x)U'_k(x)=0\, .
\end{equation}
Hence, $\hat{W}_{i0}(x)$ is independent of $k$ and we are essentially left with a linear deformation. Clearly, the range of values of $k$ can be expanded to the real numbers.

Another canonical generalization of our work is to consider deformations on the form
\begin{equation}
\hat{W}(x)=H_{0}(x){\cal F}(W)\, ,
\end{equation}
where ${\cal F}$ is any {\it functional} of the seed superpotential $W(x)$. The isospectrality condition then implies
\begin{equation}
H'_{0} (x)+(\ln {\cal F}(W))'H_{0} (x)+{\cal F}(W)H_{0} ^2(x)={\cal F}(W)^{-1}(W^2(x)+W'(x))\, .
\end{equation}
Note that $H_{0} (x)=1$ does {\it not} solve this equation unless ${\cal F}(W)=W$, since Eq.(38) with $H_0(x)=1$ implies ${\cal F}'(x)+{\cal F}^2(x)=V_+(x)$. Since $V_+(x)$ is uniquely given in Eq.(2) any other choice of functional will fail to satisfy the isospectrality condition. Hence the conclusion. The particular solution $H_0(x)=1$ is not forced upon us.  We can in principle do without it.  It is easily verified that Eq.(38) can be cast into the form Eq.(9) by the substitution
\begin{equation}
H_{0} (x)=\frac{1}{{\cal F}(W)}(\ln U(x))'\, .
\end{equation}
We can also look for an expanded solution by writing
\begin{equation}
H_{0} (x)=H_{00}(x)+\frac{1}{Z_{0} (x)}\, ,
\end{equation}
where $H_{00}(x)$ is a particular solution of Eq.(38). We then get the equation 
\begin{equation}
\frac{d}{dx}Z_{0} (x)-(\frac{d}{dx}\ln {\cal F}(W)+2H_{00}(x){\cal F}(W))Z_{0} (x)={\cal F}(W)\, ,
\end{equation}
which is a generalized form of Eq.(6). The reciprocal solution has the general form
\begin{equation}
\frac{1}{Z_{0}(x)}=\frac{e^{-2\int^x H_{00}(t){\cal F}(W)dt}}{{\cal F}(W)(C+\int^x e^{-2\int H_{00}(t){\cal F}(W)dt}du)}\, .
\end{equation}
$C$ is an integration constant, which we assume to be real. Utilizing that $H_{00}(x)={\cal F}^{-1}(x)(\ln U(x))'$ the resulting deformation coincides with Eq.(7). We thus therefore conclude that non-linear deformations on the form Eq.(37) does not generate additional deformations to the ones already generated by Eq.(1).  

\section{Deforming the Coulomb potential} 

As a relatively simple application of the linear deformation scheme let us briefly consider deformations of the Coulomb potential.  This potential has, within the framework of supersymmetric quantum mechanics, been treated in several previous works \cite{Asim}. The superpotential and the partner potential for the Coulomb potential are given by \cite{Asim}
\begin{eqnarray}
W(x)&=&\frac{q^2}{2(l+1)}-\frac{(l+1)}{x}\, ,\\
V_+(x)&=&\frac{1}{4}(\frac{q^2}{l+1})^2-\frac{q^2}{x}+\frac{(l+1)(l+2)}{x^2}\, .
\end{eqnarray}
$q$ and $l$ in these expressions are the electric charge and the angular momentum quantum numbers, respectively. These potentials result in the following general solution for $U_0(x)$ in Eq.(9) \cite{Jensen}
\begin{equation}
U_0(x)=C_1M_{l+1,l+\frac{3}{2}}(\frac{q^2x}{l+1})+C_2W_{l+1,l+\frac{3}{2}}(\frac{q^2x}{l+1})\, .
\end{equation}
The $M(x)$- and $W(x)$-functions are the Whittaker functions. The solution Eq.(10) is given by \cite{Jensen}
\begin{equation}
U_0(x)\sim e^{\frac{q^2x}{2(l+1)}-(l+1)\ln (2x)}\, .
\end{equation}
We will for simplicity assume this solution in the following. We will let $C_{01}\rightarrow\infty$ in Eq.(7) such that we deal with the identity deformation $\hat{W}_0(x)=\hat{W}_{00}(x)=W(x)$. We will also ignore the $Q_j(x)$ contributions in the following. Define $A\equiv q^2/(2(l+1))$ and $B\equiv l+1$.  It then follows that
\begin{equation}
P_1(x)=C_{11}+C_{12}\int^x t^{2B}e^{-2At}dt\, ,
\end{equation}
such that
\begin{equation}
\hat{W}_{10}(x)= A-\frac{B}{x}+\frac{C_{12}x^{2B}e^{-2Ax}}{C_{11}+C_{12}\int^x t^{2B}e^{-2At}dt}\, .
\end{equation}
Let us consider the $s$-state with $l=0$ in order to get a better grasp on the content buried in Eq.(48). We also set $q\equiv 1$.  The expression for $\hat{W}_{10}(x)$ then reduces to
\begin{equation}
\hat{W}_{10}(x)=\frac{1}{2}-\frac{1}{x}+\frac{C_{12}x^2e^{-x}}{C_{11}-C_{12}(x^2+2x+2)e^{-x}}
\end{equation}
after redefining $C_{11}$ such that the lower integration limit of the integral in Eq.(48) does not appear explicitly in the expression for the potential. We will automatically do such redefinitions in the following when it is appropriate. The corresponding physical potential $\hat{V}_{-1}(x)$ can either be derived from the definition $\hat{V}_{-1}(x)\equiv \hat{W}_{10}^2(x)-\hat{W}_{10}'(x)$ or from Eq.(11) with $\hat{W}_{00}(x)=W(x)$ and $C_{01}$ finite. This is a consequence of a regeneration of Eq.(7) by the recursion scheme which was noted in Figure 1. From the definition it follows that
\begin{eqnarray}
\hat{V}_{-1}(x)=\frac{1}{4}-\frac{1}{x}+\frac{C_{12}x(2x-4)e^{-x}}{C_{11}-C_{12}(x^2+2x+2)e^{-x}}+\frac{2C_{12}^2x^4e^{-2x}}{(C_{11}-C_{12}(x^2+2x+2)e^{-x})^2}\, . 
\end{eqnarray}
In the special case when we set $C_{11}=0$ the last term in Eq.(49) becomes independent of the exponentials (and $C_{12}$) and thus reduces to a pure rational function. The physical potential $\hat{V}_{-1}(x)$ generated by $\hat{W}_{10}(x)$ is then given by
\begin{equation}
\hat{V}_{-1}(x)=\frac{1}{4}-\frac{1}{x}+\frac{4x(x+2)}{(x^2+2x+2)^2}\equiv V_-(x)+\frac{4x(x+2)}{(x^2+2x+2)^2}\, .
\end{equation}

Let us go to the second iteration level starting from the expression for $\hat{W}_{10}(x)$ in Eq.(49) with $C_{11}=0$, for convenience. It then follows that
\begin{eqnarray}
\hat{W}_{20}(x)=\hat{W}_1(x)+\frac{C_{22}x^2e^{x}}{C_{21}(x^2+2x+2)^2+C_{22}(x^2+2x+2)e^{x}}\, .
\end{eqnarray}
Note that when $C_{21}=0$ we get $\hat{W}_{20}(x)=W(x)$. Hence, the deformation scheme allows in general for the possibility that additional iterations in particular cases may regenerate previous potentials in a nontrivial fashion. The expression for the corresponding physical potential is given by 
\begin{eqnarray}
\hat{V}_{-2}(x)=\hat{V}_{-1}(x)+\left[\frac{C_{22}x^2e^x}{C_{21}(x^2+2x+2)^2+C_{22}(x^2+2x+2)e^x}\right]\times\nonumber\\
\times\left[ -4(\frac{1}{x}+\frac{\frac{1}{2}x^2}{x^2+2x+2})+\frac{2(C_{21}(2x+2)+C_{22}e^x)}{C_{21}(x^2+2x+2)+C_{22}e^x}\right]\, .
\end{eqnarray}

The superpotential stemming from the third iteration with $\hat{W}_{20}(x)$ in Eq.(52) as the starting point is given by
\begin{equation}
\hat{W}_{30}(x)=\hat{W}_{20}(x)+\dfrac{\left(\dfrac{C_{32}x^2e^x}{C_{21}(x^2+2x+2)^2+C_{22}(x^2+2x+2)e^x}\right)}{\left( C_{31}+\dfrac{C_{32}}{C_{22}}\ln \left|\dfrac{C_{21}(x^2+2x+2)+C_{22}e^x}{x^2+2x+2}\right|\right) }\, .
\end{equation}
This superpotential introduces the possibility for a logarithmic singularity away from the origin when $C_{22}/C_{21}<0$. We note that setting $C_{31}=0$ does not regenerate a previous potential as was possible at the previous iteration level when we correspondingly put $C_{21}=0$. From Eq.(54) we can deduce the physical potential $\hat{V}_{-3}(x)$ at the third iteration level. We do not reproduce it here due to its complexity.  Due to the complicated integrals appearing we are not able to provide the analytical expression for $\hat{W}_{40}(x)$. We leave detailed studies of the Coulomb potential for the future.

\section{Conclusion}

In a previous paper we showed that isospectral deformations on the form Eq.(3) are contained in the space of deformations generated by isospectral deformations on the form Eq.(1).  In this paper we have shown that Eq.(1) can be considered as the initial, or base, deformation of a novel infinite recursive isospectral deformation chain. This thus answers to some extend the question by which we ended our previous paper \cite{Jensen}; how does the most general isospectral deformation of the kind considered there (Eq.(1) in this paper) look like.  The results in this paper do obviously only give a partial answer. We deduced in particular that a class of recursive deformations exists which is generated by the solutions of the {\it non-linear} differential equations in Eq.(16). 

We briefly discussed various ways to construct alternative recursive deformation structures.  We considered a linear product structure, polynomial deformations and completely generalized base deformations.  They all either failed to provide a recursive structure or they turned out to be identical to the linear deformation scheme.

We applied the linear recursive scheme to the Coulomb potential. We derived novel superpotentials which all satisfy the isospectrality condition. This application did also demonstrate how easily novel isospectral deformations can be generated in this approach. It did also demonstrate an increased relative complexity of the generated potentials with the number of iterations, as one also naively would expect from the expression Eq.(21).

\end{document}